\documentstyle[preprint,prd,aps]{revtex}
\begin{document}
\title{Inflation in Supersymmetric Unified Theories}
\author{R. Jeannerot\thanks{E-mail: R.Jeannerot@sussex.ac.uk}\\
{\em Centre for Theoretical Physics, University of Sussex,}\\
{\em Falmer, Brighton BN1 9QH, UK}}
\maketitle
\tightenlines
\begin{abstract}

We construct supersymmetric unified models which automatically lead
to a period of inflation. The models all involve
a {\rm U}(1) symmetry which does not belong to the MSSM. We consider
three different types of models depending on whether this 
extra {\rm U}(1) is the subgroup of a non abelian
gauge group, is a {\rm U}(1) factor belonging to the
visible sector or is a {\rm U}(1) factor belonging to the hidden sector. 
 Depending on the structure
of the unified theory, on the spontaneous
symmetry breaking pattern and on whether we have global or local
supersymmetry, inflation
may be driven by the non-vanishing vacuum expectation value 
of a F-term or by that of a D-term. 
In both scenarios cosmic strings form at the end of inflation, and
they have different properties in each model. Both inflation
and cosmic strings contribute to the CMBR temperature anisotropies. 
We show that the strings contribute to the $C_l$'s up to the
level of 75 \%. Hence the contribution from strings
to the CMBR and to the density perturbations in the early Universe
which lead to structure formation cannot be neglected. We also discuss a very
interesting class of models which involve a ${\rm U}(1)_{B-L}$ gauge symmetry.
\end{abstract}
\newpage

\section{Introduction}
\label{sec-intro}

Supersymmetric unified theories of the strong, weak and electromagnetic 
interactions,
such as the left-right model ${\rm SU}(3)_c \times {\rm SU}(2)_L \times {\rm 
SU}(2)_R \times
{\rm U}(1)_{B-L}$ or grand unified theories such as SO(10) may solve many of
the outstanding problems of both the particle physics and the
cosmological standard models \cite{SUSY}.
In particular, they can solve the question of fermion masses or lead
to automatic R-parity conservation,
 they can explain the matter anti-matter asymmetry
of the Universe and provide good dark matter candidates. 
Such theories beyond the standard model may also be
needed to explain the grand desert between the string unification scale
$\sim 5 \times 10^{17}$ GeV and the electroweak scale $\sim 10^2$ GeV, 
if string theory is the theory of quantum gravity needed to explain the
state of the Universe at the Planck scale.

In building supersymmetric grand unified models aiming to be consistent
with cosmology
one is confronted with two main problems. The first problem
is that any semi-simple grand unified gauge group 
which is broken
down to the standard model ${\rm SU}(3)_c \times {\rm SU}(2)_L \times {\rm 
U}(1)_Y$ inevitably
leads to the formation of topologically stable monopoles, according to the
Kibble mechanism \cite{Kibble}. These monopoles, if present today, would 
dominate the energy
density of the universe, and our universe would be very different
from what we observe. Even if the grand unified gauge group G
is not semi-simple, it may still be confronted with the monopole problem. 
Topologically
stable monopoles form during the spontaneous symmetry breaking (SSB) pattern 
from
$G$ down to the standard model gauge group if
the second homotopy 
group $\pi_2 ({G \over {\rm SU}(3)_c \times {\rm SU}(2)_L \times 
{\rm U}(1)_Y})$ is non trivial. For example, unified theories such as the
trinification ${\rm SU}(3)^3$ or the one based on the Pati-Salam group 
${\rm SU}(4) \times {\rm SU}(2)_L \times {\rm SU}(2)_R$ also lead to the 
formation of topologically stable 
monopoles making the models inconsistent with observations on their own. Hence 
some
mechanism has to be invoked to get rid of these monopoles. The mechanism
which is the most promising is that of inflation. Because inflation not only 
solves
the monopole problem, but also predicts the formation of large scale structures,
it predicts anisotropies 
in the temperature fluctuations of the cosmic microwave background radiation
(CMBR), and it solves the horizon
problem \cite{inflation}. The second problem, which is directly related to this 
first one,
is that inflation usually requires very fine tuning of the parameters. 

For the monopole problem to be cured, the period
of inflation must take place after the phase transition leading to the
formation of the unwanted monopoles, and after any other phase transition
associated with the formation of other catastrophic defects such as
domain walls. Once a scenario for inflation has
been chosen, the study of topological defect formation before and after 
inflation and
the various possible SSB patterns from the unified gauge group
down the the standard model, together with the requirement that a scenario for 
baryogenesis 
occurs after inflation, allows one to select the SSB patterns which are 
cosmologically 
interesting. For example, SSB pattern selection from topological defect 
formation in a supersymmetric SO(10) models can be found in \cite{paper2}.

If inflation
comes from the (grand) unified theory (GUT) beyond the standard model which 
describes the symmetries between particles at $\sim 2 \times 10^{16}$ GeV, it
is probably of the hybrid type \cite{hybrid}. This is because hybrid
inflationary scenarios occur without fine-tuning and arise in a {\em natural}
way. By {\em natural} we mean that, in the models considered below, 
no extra field is needed, apart from a scalar field singlet
under the considered gauge group. This extra field is
probably anyway needed to build the model to force the Higss field 
used to lower the rank of the group to acquire a 
vacuum expectation value. We point out that potentials 
which do not require the existence of the scalar singlet can be used
to lower the rank of the group, but then inflation cannot arise from this 
sector since it does not involve any dynamical field. Note that inflation 
could also come from string theory for example. We do not consider this
case here. 

There are two main possibilities for the hybrid inflationary scenario to 
be implemented. The period of inflation can arise from the GUT considered 
itself,
by for example coupling some GUT Higgs fields to a scalar field singlet under
the considered gauge group which plays the role of the inflaton, or it can come 
from
another sector, part of the visible sector which must then involve some extra 
gauge or global symmetry, or from the hidden sector. Inflation 
assumes that there was a period in the very early
universe when the potential, vacuum energy density was dominating the energy 
density of the universe, so that the cosmic scale factor grew exponentially.
Since in supersymmetric theories the scalar potential is basically the sum of 
the F-terms and D-terms,
inflation may be driven by either the non-vanishing vacuum 
expectation value (vev) of a F-term or by 
that of a D-term. 
Almost
all inflationary scenarios in supersymmetric theories considered in the 
literature rely on the non-vanishing vev of a 
F-term. This is, however, usually a problem 
in supergravity theories \cite{Cop}. Indeed, for inflation to occur and lead to 
a nearly scale 
independent spectrum of density perturbations, as suggested by observations, 
the potential must be 
flat in the inflaton direction, and the inflaton must have a very small mass.
During inflation, the energy density of the universe is dominated by
the vacuum energy density, which is usually close to 
$V_0^{1\over 4} \sim M_{GUT}$, and all the
scalars of the theory acquire a mass $m^2 = {V_0 \over M^2_{pl}}$.
The inflaton mass is then much greater than the Hubble constant H
and the slow roll conditions, which require ${V" \over V} << H$,
where primes denote derivative with respect to the inflaton field, cannot be 
satisfied. A way out of that problem 
is to consider special form of the superpotential as well as particular
initial conditions \cite{RiottoLinde}, special form of the K\"ahler potential
\cite{Cop,Stewart}, or extra symmetries
such as a global {\rm U}(1) \cite{Freese}. We do not consider these cases here.
Now when $V_0$ is the result of a non-vanishing D-term, scalar fields which 
are uncharged under the 
unified gauge group, such as the inflaton field, only get masses 
$m \ll H$ \cite{Dvali}. The slow conditions are satisfied and inflation can take
place. Hence in supergravity, hybrid inflationary scenarios driven by
the non-vanishing vev of a D-term are favoured.

D-term inflation requires the existence of a Fayet-Illiopoulos term \cite{FI}, 
which can only exist is the group contains a {\rm U}(1) factor with $TrQ \neq 0$ 
(where Q is the {\rm U}(1) charge) \cite{Dterm}. Hence the Fayet-Iliopoulos 
D-term of a {\rm U}(1) gauge symmetry cannot be generated in any order of 
perturbation 
theory if this {\rm U}(1) is at some arbitrary scale unified in a non abelian
gauge group. Thus D-term inflation can arise in the visible 
sector or in the hidden sector only if there is a {\rm U}(1) factor with $TrQ 
\neq 0$
in the appropriate sector. (Though dynamical generation of a 
D-term at an intermediate symmetry breaking scale may be possible, we do 
not consider this as a possible mechanism for inflation here).
 Hence for a semi-simple gauge group, 
if inflation arises from the GUT itself, it can only be driven by the 
non-vanishing vev of a F-term. 

The plan of this paper is as follows:

In Sec.\ref{sec-setup} we briefly review the idea of supersymmetric 
unified theories.

In Sec.\ref{sec-building} we describe an easy and useful way of building a 
unified theory 
of the strong,
weak and electromagnetic interactions which automatically lead to 
 a period of false vacuum
hybrid inflation \cite{Cop}. `Hybrid' because the inflaton field couples to some 
Higgs
fields which are
used to break a {\rm U}(1) gauge symmetry, and `false vacuum' because during 
inflation, which occurs for
values of the inflaton field much greater than a given critical value, the 
Higgs fields are trapped at a local minimum of the potential. We
consider both F-term inflation with the simplest possible 
superpotential \cite{Cop,DvaSha} and D-term
inflation \cite{Pierre,Halyo}. The models which are discussed are
all based on rank greater than five theories. They all 
involve a {\rm U}(1) gauge symmetry which does not belong to the minimal
supersymmetric standard model (MSSM).
We discuss three main class of models, depending on
whether this extra {\rm U}(1) (i) is the subgroup of a non abelian gauge group 
(ii) is a {\rm U}(1) factor belonging to the
visible sector or (iii) is a {\rm U}(1) factor belonging to the hidden sector.

All the models which are discussed in this paper lead to the formation of 
cosmic strings 
at the end of inflation. 
The formation
of cosmic strings at the end of inflation is discussed in Sec.\ref{sec-strings}.
Note that cosmic strings formation at the end of inflation in a somewhat 
different 
context has been discussed before \cite{ShafiLazcs}.
In Sec.\ref{sec-infl}, we determine the spectral index
of density perturbations coming from inflation and the
inflationary scale. We also determine the relative contributions from inflation 
and cosmic strings to the CMBR.

In Sec.\ref{sec-models}, we discuss in details the three main classes of models 
to which the construction described in Sec.\ref{sec-building} applies. 
We illustrate each case with and interesting class of unified models: 
theories beyond the standard model
which contain a ${\rm U}(1)_{B-L}$ gauge symmetry.

In Sec.\ref{sec-concl} we summarise our main results and conclude.

\section{The setup}
\label{sec-setup}

In studying supersymmetric unified theories, the picture we have in mind is
that the very early universe underwent a series of phase transitions associated
with the SSB pattern:
\begin{eqnarray}
G  \times{\rm SUSY} \stackrel{M_{GUT}}\rightarrow ...\rightarrow H \times{\rm 
SUSY}
&\rightarrow & {\rm SU}(3)_c \times {\rm SU}(2)_L 
\times {\rm U}(1)_Y (\times{\rm SUSY}) \nonumber \\
 &\stackrel{M_{\rm Z}}\rightarrow & {\rm SU}(3)_c \times {\rm U}(1)_Q 
\label{eq:SSB}
\end{eqnarray}
where G is the (grand) unified gauge group (not necessarily semi-simple) 
and{\rm SUSY} stands for supersymmetry. 
The unified scale is $M_{GUT} \sim 2 \times 10^{16}$ GeV,  the 
electroweak scale is $M_{Z} \sim 100$ GeV and supersymmetry
is broken at $\sim 10^3$ GeV. Then there also may or may not be an underlying
theory such as superstrings valid at higher energies. 
 If this theory, which must be a theory of quantum gravity, is
superstring theory, then the grand unified gauge coupling constant $g_G$
would run up to the string scale $M_{string} \sim 5 \times 10^{17}$ GeV
where it would then unify with the gauge coupling of the hidden gauge 
group and with the gravitational coupling. 

If G is assumed to be recovered
from superstring compactification schemes, there will be some additional 
symmetries
in both the visible sector
and the hidden sector (though not necessarily in the visible sector). 
In this paper, we suppose the existence of a {\rm U}(1)
symmetry, namely ${\rm U}(1)_x$, which does not belong to the minimal
supersymmetric standard model and can either be a subgroup of 
G or a {\rm U}(1) factor belonging to the visible sector or a {\rm U}(1) factor
belonging to the hidden sector in the appropriate case. 
If G is assumed to be recovered
from superstring compactification schemes, then 
there is a good reason for a {\rm U}(1) factor to be standing there,
since extra {\rm U}(1) factors normally appear in effective field theories
arising from strings.
The low energy spectrum will also be constrained, in particular some
high dimensional Higgs representations of the unified theory G may not be 
allowed 
\cite{string GUTs}, such as the 126 dimensional Higgs representation 
in the case
of SO(10) \cite{string126}. This will influence the choice of the
grand unified gauge group, the SSB pattern and the choice
of the inflationary scenario. 

But supersymmetric unified theories can also be thought
as standing there on their own. This gives
more freedom in choosing the SSB pattern and the 
Higgs representations to use to implement it, the inflationary scenario and 
the way of getting some desirable phenomenological effects such as neutrino 
masses and R parity conservation.

\section{Building a model}
\label{sec-building}

We consider here an easy and useful way of building a supersymmetric 
unified model which gives rise to a false vacuum hybrid inflationary scenario 
without fine tuning, which is independent
of the process of supersymmetry breaking at low energy, and independent
of the form of the supersymmetry breaking parameters.

\subsection{Assumptions}
\label{sec-assume}

We suppose the existence of a ${\rm U}(1)_x$ gauge symmetry which does not 
belong
to the MSSM and spontaneously breaks at a scale $M_x$. 
The ${\rm U}(1)_x$ is related to the
inflationary scenario: the inflaton field couples with a pair of Higgs field
$\Phi_x$ and $\overline{\Phi}_x$
which are used to break ${\rm U}(1)_{x}$. 
The inflationary scenario may be either of the F-term hybrid type
with the simplest possible superpotential \cite{Cop,DvaSha} or 
D-term hybrid type \cite{Pierre,Halyo}, depending
on whether the supersymmetry is global or local, on the 
SSB pattern and on the origin of ${\rm U}(1)_x$.
We consider three distinct classes of models, depending on whether the
${\rm U}(1)_x$  (i)
is  a subgroup of the considered grand unified gauge group G, and
the SSB given by:
\begin{eqnarray}
G \times{\rm SUSY} &\stackrel{M_{GUT}}\rightarrow & ... \rightarrow (H \supset 
{\rm U}(1)_x)
\times{\rm SUSY} \stackrel{M_{\rm x}}\rightarrow 
(K \not\supset {\rm U}(1)_x )\times{\rm SUSY} \rightarrow ... \nonumber \\
&\rightarrow & {\rm SU}(3)_c \times {\rm SU}(2)_L 
\times {\rm U}(1)_Y \times{\rm SUSY} \stackrel{M_{\rm Z}}\rightarrow {\rm 
SU}(3)_c \times {\rm U}(1)_Q ,
\end{eqnarray}
(ii) is an extra {\rm U}(1) symmetry belonging to the visible sector without 
being
a subgroup of a non abelian gauge group, and the SSB is given by:
\begin{eqnarray}
G \times {\rm U}(1)_{x} \times{\rm SUSY} & \stackrel{M_{GUT}}\rightarrow& ...
\rightarrow 
H \times {\rm U}(1)_{x} \times{\rm SUSY} \nonumber \\
&&\stackrel{M_{x}}\rightarrow 
 {\rm SU}(3)_c \times {\rm SU}(2)_L 
\times {\rm U}(1)_Y (\times{\rm SUSY}) \nonumber \\
 &&\stackrel{M_{\rm Z}}\rightarrow {\rm SU}(3)_c \times {\rm U}(1)_Q 
\end{eqnarray}
(iii) is an extra {\rm U}(1) symmetry 
belonging to the hidden sector which is not the subgroup of a non abelian gauge 
group, and the SSB then given by:
\begin{eqnarray}
G \times [ {\rm U}(1)_{x} ]_{hidden} \times{\rm SUSY} 
&\stackrel{M_{GUT}}\rightarrow & ...
H \times [ {\rm U}(1)_{x} ]_{hidden} \times{\rm SUSY} \nonumber \\
&&\stackrel{M_{x}}\rightarrow  {\rm SU}(3)_c \times {\rm SU}(2)_L 
\times {\rm U}(1)_Y (\times{\rm SUSY}) \nonumber \\
 &&\stackrel{M_{\rm Z}}\rightarrow {\rm SU}(3)_c \times {\rm U}(1)_Q .
\end{eqnarray}
In each case, the scale $M_x$ will be constraint by the four year COBE DMR data, 
see Sec.\ref{sec-infl}.

\subsection{The superpotential}
\label{sec-super}

The superpotential which will implement the full SSB
pattern from $G (\times {\rm U}(1)_x)$ down to ${\rm SU}(3) \times {\rm U}(1)_Q$ 
and gives
rise to a period of inflation is constructed 
by considering different sectors, and by adding the superpotentials
describing each sector. 

The different sectors are:
\begin{itemize} 

\item The inflaton sector, which 
breaks
the ${\rm U}(1)_x$ gauge symmetry mentioned above, it is described by 
$W_{infl}$.

\item The GUT sector which implements
the SSB of $G$ down to 
the standard model gauge group 
(apart from that of ${\rm U}(1)_x$ when applicable), it is described 
by $W_{GUT}$.

\item The electroweak sector which breaks ${\rm SU}(3)_c \times {\rm SU}(2)_L 
\times {\rm U}(1)_Y$
down to ${\rm SU}(3)_c \times {\rm U}(1)_Q$ and is described by $W_{ew}$.

\item And a hidden sector (which also includes the ${\rm U}(1)_x$
when appropriate) which is described by 
$W'  _{hidden}$. Note that in supergravity, by hidden sector we mean a sector of 
the
theory that couples to the observable sector of quarks, leptons, gauge fields,
higgses and their supersymmetric partners only through gravitational 
interactions. Since globally supersymmetric theories do not include
gravity, it probably does not really make sense to talk about
hidden sector in that case. Hence in globally supersymmetric theories we
distinguish the visible sector from the hidden sector by assuming
that the $\Phi_x$ and $\overline{\Phi}_x$ Higgs fields  used to break ${\rm 
U}(1)_x$ 
carry G quantum numbers or not and if the particle contain of the MSSM
carries ${\rm U}(1)_x$ quantum numbers or not.

\end{itemize}
Note that some of these sectors will still be linked for doublet-triplet
splitting purpose or the avoidance of unwanted massless pseudo-Goldstone
bosons. If they are linked, it
must be done in such a way that it does not destabilise the required
vevs. Additional link terms (i.e. not actually necessary to implement 
the SSB pattern nor the inflationary scenario), must be included in
$W_{GUT}$ and $W_{ew}$. 

The resulting superpotential is then written as
\begin{equation}
W_{tot} = W_{GUT}(A's) + W_{infl}(S,\Phi_{x},\overline{\Phi}_{x}) + 
W_{ew}(H_1,H_2) + W'_{hidden}(Z's)
\label{eq:super}
\end{equation}
where the $A$'s, $H_1$ and $H_2$ are Higgs superfields with the bosonic 
components in appropriate representations of $G$. In this paper, we will
use the same notation for the chiral superfields and their bosonic
components if there is no risk of confusion.
$H_1$, and $H_2$ must contain {\rm SU}(2) doublets. 
S is a scalar field singlet
under $G$ which plays the role of the inflaton. $\Phi_{x}$ and
$\overline{\Phi}_{x}$ 
are Higgs superfields used to break the 
${\rm U}(1)_x$ gauge symmetry. They have opposite $x$ charges. 
They may transform non-trivially
under $G$ where appropriate, see Sec.\ref{sec-models}. 
 $Z's$ are Higgs superfields which belong
to the hidden sector and interact only gravitationally with the
other fields. 

Note that if we use the general splitting
of $W_{tot}$ into the two main sectors which are the visible and 
the hidden sector \cite{Dine}:
\begin{equation}
W_{tot} = W_{visible} + W_{hidden},
\end{equation}
we have $W_{visible} = W_{GUT} + W_{infl} + W_{ew}$,
$W_{hidden} = W'_{hidden}$ in cases (i) and (ii) and 
$W_{visible} = W_{GUT} + W_{ew}$,
$W_{hidden} = W_{infl} + W'_{hidden}$ in case (iii).

The superpotential in the inflaton sector will look different whether inflation
is driven by the non-vanishing vev of a F-term or by that of a D-term. 
Extra symmetries must be imposed for each particular case. 
The choice for F-term or D-term inflation is model dependent, see 
Sec.\ref{sec-models}.

In this paper, we consider
the simplest superpotentials which lead to a period of F-term and D-term 
inflation respectively, and do not involve any non-renormalisable
term, but generalisation to inflationary scenarios with
more complicated superpotentials in the inflaton sector is straight forward.

In the case of F-term inflation, the
superpotential in the inflaton sector is chosen to be\cite{Cop,DvaSha}:
\begin{equation}
W_{infl}(S,\Phi_x,\overline{\Phi}_x) = \alpha S \Phi_x \overline{\Phi}_x - 
\mu^2 S .
\label{eq:Finfl}
\end{equation}
The $\Phi_x$ and $\overline{\Phi}_{x}$ are Higgs fields used to break ${\rm 
U}(1)_x$
in appropriate representations. $S$ is a scalar superfield singlet 
under G which plays the role 
of the inflaton field. $\alpha$ and $\mu$ are two positive constants,
and the ratio ${\mu \over \sqrt{\alpha}}$ sets the ${\rm U}(1)_x$ symmetry 
breaking scale
$M_x$. The superpotential given by Eq.(\ref{eq:Finfl}) is the most 
general potential consistent with a continuous R symmetry under 
which the fields transform as $S \rightarrow e^{i\gamma} S$, 
$\Phi_x \rightarrow e^{i\gamma} \Phi_x$,
$\overline{\Phi}_x \rightarrow e^{i\gamma} \overline{\Phi}_x$ and 
$W \rightarrow e^{i\gamma} W$. Note that this superpotential with the assumption 
of 
minimal K\"ahler potential in supergravity is viable \cite{RiottoLinde}.

If inflation comes from the non-vanishing vev of a D-term, this can only be the 
case
when the ${\rm U}(1)_x$ 
gauge symmetry is not the subgroup of a non-abelian gauge group, the
superpotential in the inflaton sector is \cite{Pierre,Halyo}:
\begin{equation}
W_{infl}(S,\Phi_x,\overline{\Phi}_x) = \alpha S \Phi_x \overline{\Phi}_x 
\label{eq:Dinfl}
\end{equation}
$\Phi_x$ and $\overline{\Phi}_x$ are Higgs fields 
with $x$-charges $+Q$ and $-Q$ respectively, which break ${\rm U}(1)_x$
when acquiring vevs. In the following, we
will set $Q=1$. Note that the qualitative results given in this paper
will be unchanged for higher charged Higgs fields.
$S$ is a scalar superfield singlet under $G$ which plays the role 
of the inflaton, and $\alpha$ is a positive constant. 
The potential given by Eq.(\ref{eq:Dinfl}) is the most 
general potential consistent with a continuous R symmetry under 
which $S \rightarrow e^{i\gamma} S$, $\Phi_x \rightarrow e^{i\gamma} \Phi_x$,
$\overline{\Phi}_x \rightarrow e^{i\gamma} \overline{\Phi}_x$ and 
$W \rightarrow e^{i\gamma} W$. There is also a $Z_3$ symmetry under which
the fields transform as $\chi \rightarrow e^{i{2\pi \over 3}} \chi$, for 
$\chi=S, \Phi_x$ and $\overline{\Phi}_x$. The existence of a Fayet-Iliopoulos 
D-term for the ${\rm U}(1)_x$
symmetry is assumed. It sets the ${\rm U}(1)_x$ symmetry breaking scale
as well as the inflationary scale, see Sec.\ref{sec-infl}.

\subsection{The scalar potential}
\label{sec-scalar}

\subsubsection{Globally supersymmetric case}
\label{sec-global}

In globally supersymmetric theories, the scalar potential 
$V_{tot}$ which can be derived from the superpotential given in
Eq.(\ref{eq:super}) is \cite{SUSY}:
\begin{equation}
V_{tot} = \sum_i |F_{i}|^2 + {1\over 2} \sum_\alpha |D_\alpha|^2 
+ V_{soft} \label{eq:Vtot}
\end{equation}
where $i$ runs from 1 to N, where N is the number of chiral superfields 
in $ W_{tot}$. $V_{soft}$ contains all the soft terms generated by 
supersymmetry breaking at low energy. The F and 
D-terms are respectively given by: 
\begin{equation}
F^i = {\partial W \over \partial h_i } \label{eq:Fglob}
\end{equation}
and
\begin{equation}
D^\alpha = g \sum_{a,b} h^*_a T^{\alpha a}_{b} h^b  + \xi_\alpha 
\label{eq:Dglob}
\end{equation}
where the $h_i$ are the scalar components of the chiral superfields
included in the superpotential $W_{tot}$ and $T^\alpha$ are the generators of 
$G (\times {\rm U}(1))$.  $\xi_\alpha$ is a Fayet-Illiopoulos term which can 
only exist when $T^\alpha$ is the generator of the {\rm U}(1) factor. The 
conditions 
for unbroken global supersymmetry is that all the F and D-terms equal zero.

From Eqs. (\ref{eq:super}), (\ref{eq:Vtot}), (\ref{eq:Fglob}) 
and (\ref{eq:super}), we find that the full scalar potential
can be written as:
\begin{equation}
V_{tot} = V_{GUT} + V_{infl} + {\it V_{ew} + V_{soft}} . \label{eq:scalar}
\end{equation}
$V_{infl}$ has a global minimum such that the ${\rm U}(1)_x$ symmetry  
is broken down to unity (or eventually a discrete $Z_n$). 
$V_{GUT} (+ V_{infl})$ (when ${\rm U}(1)_x$ is a subgroup of G)
has a global 
minimum such that the gauge group G is broken down to 
${\rm SU}(3)_c \times {\rm SU}(2)_L \times {\rm U}(1)_Y$
and supersymmetry is unbroken. Note that when G is a semi-simple gauge group
and ${\rm U}(1)_x$ is a subgroup of G, the 
$\Phi_x$ and $\overline{\Phi}_x$, which transform non trivially under G, 
break ${\rm U}(1)_x$ as well some other generators of G. If some generators of G 
are broken in both the inflaton and the GUT sector, this will result in
the existence of pseudo-Goldstone bosons which may be undesirable. The problem
can be cured by linking the two sectors. As mentioned above, see 
Sec.\ref{sec-super}, this
must be done in such a way that it does not destabilise the required vevs. 
This is not
always easy \cite{babu}. $V_{GUT} (+ V_{infl}) + V_{ew}$  has a global 
minimum such that G is broken down to ${\rm SU}(3)_c \times {\rm U}(1)_Q$. The 
last two 
terms 
in Eq.(\ref{eq:scalar}) should not affect the behaviour of the fields in the GUT 
and
in the inflationary sector at high energies. Indeed, as pointed out
by Dvali \cite{Dvali}, in the very early universe, at temperatures of
order $10^{15}-10^{16}$ GeV, the fields do not know whether supersymmetry will
be broken at lower energy or not. Hence, the dynamics of the scenario should
be independent of the soft masses for the supersymmetric particles at low 
energy.  

We know briefly discuss the scalar potentials for both F-term and D-term 
inflation, and
the dynamics of the fields in the inflaton sector. They have been
discussed before \cite{DvaSha,Pierre}. Discussion concerning
the evolution of all Higgs fields can be found in Sec.\ref{sec-models} for
each appropriate model.

When inflation is driven by the non-vanishing vev of a F-term, 
the superpotential in the inflaton sector is given by 
Eq.(\ref{eq:Finfl}), and the Fayet-Iliopoulos D-term for ${\rm U}(1)_x$ 
vanishes. The 
scalar potential in the inflaton sector is then given by:
\begin{equation}
V^F_{infl} = \alpha^2 |S|^2 ( |\Phi_x|^2 + |\overline{\Phi}_x|^2 ) +
|\alpha \overline{\Phi} \Phi - \mu^2|^2  . \label{eq:Fscalar}
\end{equation}
$V^F_{infl}$ has a unique supersymmetric minimum corresponding to  
$\langle |\Phi | \rangle = \langle | \overline{\Phi} | \rangle = {\mu \over 
\sqrt{\alpha }}$
and $S = 0$. It also has a local minimum for $|S| > {\mu \over \sqrt{\alpha}}= 
S_c$, 
at $\langle |\Phi_x |\rangle =  \langle | \overline{\Phi}_x | \rangle = 0$. 
${\mu \over \sqrt{\alpha }}$ sets the ${\rm U}(1)_x$ symmetry breaking scale 
$M_x$. Let's assume
chaotic initial conditions. The potential is very flat in the $|S|$ direction,
and the $\Phi_x$ and $\overline{\Phi}_x$ fields settle down to the
local minimum of the potential, $\Phi_x = \overline{\Phi}_x = 0$ . 
The universe is dominated by a non
vanishing vacuum energy density, $V_0^{1\over 4} = \mu$, inflation starts, 
and supersymmetry
is broken. There are therefore some quantum corrections to the effective 
potential \cite{DvaSha}:
\begin{equation}
V^F_{eff} = \mu^4 (1 + {\alpha^2 \over 16 \pi^2} \ln{\alpha^2 
|S|^2 \over \Lambda^2})
\end{equation}
where $\Lambda$ is a renormalisation constant. These corrections
help the inflaton field to slowly roll down the potential. When  $|S|$ 
falls below $S_c$ and the 
$\Phi_x$ and $\overline{\Phi}_x$ fields quickly 
settle down to the global minimum of the potential 
spontaneously breaking ${\rm U}(1)_x$. Supersymmetry is restored.

For D-term inflation, the superpotential in the inflation
sector is given by Eq.(\ref{eq:Dinfl}), and the existence of a Fayet-Iliopoulos 
D-term for the ${\rm U}(1)_x$ gauge
symmetry, namely $\xi_x$, is assumed. It will set the ${\rm U}(1)_x$ symmetry 
breaking scale $M_x$
as well as the inflationary scale, see Sec.\ref{sec-infl}. 
The scalar potential in the inflaton sector
is now given by \cite{Pierre,Halyo}:
\begin{equation}
V^D_{infl} = \alpha^2 |S|^2 (|\Phi_x|^2 + |\overline{\Phi}_x|^2 ) +
\alpha^2 |\overline{\Phi}_x\Phi_x|^2 + {g^2 \over 2} (|\overline{\Phi}_x|^2
-|\Phi_x|^2 + \xi_x)^2 . \label{eq:Dscalar}
\end{equation}
To summarise,
$V_{infl}$ has a unique supersymmetric minimum where 
\begin{equation}
\langle S\rangle = 
\langle \overline\Phi \rangle = 0 \, \, {\rm and} \, \,  \langle |\Phi|\rangle = 
\xi_x^{1\over 2}. \label{eq:min}
\end{equation}
It also has a local minimum for $|S| > S_c = {g\over \alpha} \xi_x^{1\over 2}$
at $\langle \Phi_x \rangle = \langle \overline{\Phi}_x \rangle = 0$, and $V = 
{g^2 \over 2} \xi_x^2$.
Hence setting chaotic initial conditions, $|S|$ is initially much greater
than $S_c$ and the  $\Phi_x$ and $\overline{\Phi}_x$ fields 
settle down the local minimum. Inflation can take place
and supersymmetry is broken. This leads to a one loop correction to 
the potential (\ref{eq:Fscalar}) which is then given by \cite{Pierre,Halyo} :
\begin{equation}
V^D_{eff} = {g^2 \over 2} \xi_x^2 (1 + {g^2 \over 16 \pi^2} \ln{\alpha^2 
|S|^2 \over \Lambda^2}) \label{eq:Dscalareff}
\end{equation}
where $\Lambda$ is a renormalisation scale. This loop correction drives the 
inflaton 
slowly down the potential, independently of the process of supersymmetry 
breaking at low energy. The slow roll conditions are satisfied for
 $|S|^2 >> {M_{pl} g\over 8 \pi^{3\over 2}} = S_{end}$. 
 When $|S|$ falls below $S_{end}$ inflation 
ends. When $|S|$ falls
below $S_c$, the $\Phi$ and $\overline{\Phi}$ fields quickly reach the
global minimum of the potential. What is important is that the scenario does 
not depend on the process 
of supersymmetry
breaking at low energy. The value of the Fayet-Iliopoulos constant $\xi_x$ 
will be determined for successful inflation.

\subsubsection{Locally supersymmetric case}
\label{sec-local}

In supergravity theories, the full superpotential which implements 
the SSB pattern, the inflationary scenario 
and the breaking of susy can also be written as in Eq.(\ref{eq:super}).
 Note that the 
superpotential in each sector
will in general be different from the one used in the global case. 

The scalar potential in supergravity theories is given by \cite{SUSY}:
\begin{equation}
V = e^{K \over M^2} [ (K^{-1})_i^j F_i F^j - 3 {|W|^2 \over M^2}] + 
{g^2 \over 2} {\rm Re}[f_{\alpha\beta}] D^\alpha D^\beta
\end{equation}
where $M= {M_{pl} \over \sqrt{8 \pi }}$ is the reduced Planck mass. 
$K(H, \overline{H})$ is the K\"ahler function, here $H$ denotes
any Higgs superfield in the appropriate representation, or a scalar superfield, 
$W(H)$ is the superpotential,
and $f(H)$ is the gauge kinetic function. Upper (lower) indexes $(i,j)$ denote
derivatives with respect to the scalar components 
of the corresponding chiral superfields $h_i$ ($h^*_i$). The F-terms 
in supergravity theories differ from those in the globally 
supersymmetric case and are given by
\begin{equation}
F^i = {\partial W \over \partial h_i} + {\partial K \over \partial h_i} 
{W\over M} \label{eq:Fsugra}
\end{equation}
 The supergravity D-terms are
\begin{equation}
D^\alpha = K^i (T^\alpha )_i^j h_j + \xi_\alpha
\end{equation}
where $T^\alpha$ are the generators of the appropriate gauge group in the
corresponding representation and $\xi_\alpha$ is a Fayet-Illiopoulos term
which can only exist when $T^\alpha$ is the generator a abelian
group with $Tr Q \neq 0$.
It is easy to check that when $M \rightarrow \infty $, the global supersymmetric
potential is recovered.

The conditions for unbroken supergravity is that the supergravity F and D 
terms
vanish. Hence in locally supersymmetric unified theories 
the superpotential in each sector will in general be
different from the one used in the global case. The second reason
for these superpotentials to be different is that most of 
the superpotentials used to build globally supersymmetric 
GUTs do not vanish and consequently lead to a non-vanishing cosmological 
constant.
To have get a vanishing cosmological constant the superpotential in each
sector should vanish. For example, Higgs superpotential
used to build globally supersymmetric GUTs usually satisfies 
$\langle W_{GUT}\rangle \sim M_{GUT}^3$ which would give
rise to an enormous negative cosmological constant, inconsistent with
observations. The cosmological constant problem can be cured,
for example, by adding a term to the superpotential such that $\langle 
W_{GUT}\rangle = 0$. 

In what follows, we assume minimal kinetic terms $f^{\alpha\beta} = 1$ and
minimal K\"ahler potential $K = \sum_i H_i^* H^i$.
We also assume that  $\langle W_{GUT}(A)\rangle = 0$ and 
$\langle W_{infl}(S,\Phi_x,\overline{\Phi_x})\rangle = 0$, for $A, S, \Phi_x$ 
and
$\overline{\Phi}_x$ acquiring vev such that G is broken down to
${\rm SU}(3)_c \times {\rm SU}(2)_L \times {\rm U}(1)_Y$ and ${\rm U}(1)_x$ is 
broken down to unity (note that in case (i) ${\rm U}(1)_x$ is 
a subgroup of the grand unified gauge group G). 
We also assume that the 
$\overline{\Phi}_x$ is initially set to zero. 

Whereas in globally supersymmetric GUTs both F-term and D-term inflation
can occur, in supergravity GUTs D-term inflation is strongly favoured,
see Sec.\ref{sec-intro}. This requires the ${\rm U}(1)_x$ gauge symmetry to be a 
{\rm U}(1)
factor with a non-vanishing Fayet-Iliopoulos D-term. We consider this
case here.

The GUT symmetry breaking at high energy 
and the inflationary scenarios are independent of the supersymmetry breaking
mechanism at low energy, as was justified earlier. With the assumption 
of minimal supergravity, the scalar potential at high energies is:
\begin{equation}
V = e^{K\over M^2} [ F_{GUT}F^{GUT} + F_{infl}F^{infl} - 
3 {(W_{GUT} + W_{infl})^2 \over M^2} ] + {g^2 \over 2} D^{GUT}D_{GUT} + 
{g^2 \over 2} D^{infl}D_{infl}
\end{equation}
where $F_{GUT}$, $F_{infl}$, $D_{GUT}$ and $D_{infl}$ are the sum of the 
F and D-terms
for the fields in the GUT and the inflation sector respectively.
Recall also that the $A, S, \Phi_x$ and
$\overline{\Phi}_x$ fields must satisfy the vanishing conditions of the 
F and D-terms for unbroken supergravity and hence $ \langle F_{GUT} \rangle = 
0$, 
$\langle F_{infl} \rangle = 0$, $\langle D_{GUT} \rangle = 0$ and $\langle 
D_{infl} \rangle = 0$. 
With the above   choice (tuning)
of the parameters, the cosmological constant at high energy vanishes before
and after the inflationary period. 

The scalar potential now reads:
\begin{eqnarray}
V &=& e^{\sum_i |A_i| + |S|^2 + |\Phi_x|^2 + |\overline{\Phi}_x|^2 \over M^2} 
[|\Phi_x\overline{\Phi}_x|^2 ( 1 +{ |S|^4\over M^2}) + |S \overline{\Phi}_x|^2 
( 1 + {|\Phi_x|^4\over M^2}) \\
&& + |S \Phi_x|^2 ( 1 + {|\overline{\Phi}_x|^4\over M^2})
+ 3 {|S \Phi_x \overline{\Phi}_x|^2 \over M^2} ] + 
{g^2 \over 2} (|\overline{\Phi}_x|^2 -|\Phi_x|^2 + \xi_x)^2 .
\end{eqnarray}
As in the globally supersymmetric case, the potential as a local minimum 
for $|S| > S_c = {g \over \alpha} \xi_x^{1\over 2}$ and  $\Phi_x = 
\overline{\Phi}_x = 0$ and a locally supersymmetric one with 
$S = \overline{\Phi}_x =0$
and $|\Phi_x |=\xi_x^{1\over 2}$ and vanishing cosmological constant. The 
behaviour
of the fields is then similar to the globally supersymmetric case. 

\section{Constraints from COBE data}
\label{sec-COBE}

In this section, we determine the scale of inflation and 
the relative contributions from inflation 
and cosmic strings to the CMBR. We mainly focus on the case of
D-term inflation, where string formation
at the end of inflation has not yet been discussed.

\subsection{Cosmic strings form at the end of inflation}
\label{sec-strings}

Recall that cosmic strings are one dimensional topological defects
which form according to the Kibble mechanism \cite{Kibble} 
at the phase transition associated
with the SSB of a group $G$ down to a subgroup
$H$ of $G$ if the vacuum manifold $G\over H$ contains 
non contractible loops. In other words, cosmic strings form
when $ G \rightarrow H$ if the first homotopy group
$\pi_1({G\over H})$ is non trivial. For review on cosmic strings the reader is 
referred to Refs. \cite{ShelVil,Mark}. 

The simplest well know example 
of SSB which leads to the formation of cosmic strings is that of the 
abelian Higgs model, when a {\rm U}(1) symmetry breaks down to the
identity. Since in both the F-term and D-term inflationary scenarios discussed 
in this paper, the
inflaton field couples to a pair of Higgs fields used to break a ${\rm U}(1)$
gauge symmetry when acquiring vev at the end of inflation, cosmic strings 
are expected to form. It is indeed
easy to check that the scalar
potentials in the inflationary sector both the F-term and D-term inflationary
scenarios, given by Eqs.(\ref{eq:Fscalar}) and 
(\ref{eq:Dscalar}) respectively, have string solutions.
For F-term inflation, the scalar potential is given by
(\ref{eq:Fscalar}), both $\Phi_x$ and $\overline{\Phi}_x$ acquire a 
non-vanishing vev, $\langle|\Phi_x|\rangle = \langle |\overline{\Phi}_x|\rangle 
= 
{\mu \over \sqrt{\alpha}}$, and the potential is minimised for
$\arg (\Phi_x) + \arg (\overline{\Phi}_x) = 0$. The two fields
$\Phi_x$ and $\overline{\Phi}_x$ conspire to form the gauge string 
\cite{Morris}. Now in the D-term inflationary scenario the scalar potential
is given by Eq. (\ref{eq:Dscalar}), only the
$\Phi_x$ field acquires a non-vanishing, the string is a Nielsen-Olesen 
string \cite{ShelVil,Mark} and the Higgs field forming the string is $\Phi_x$.
Therefore cosmic strings
form at the end of inflation in both the F-term and D-term inflationary 
scenarios. Whether these strings are topologically stable is model
dependent. They are stable in models (ii) and (iii), and almost always
stable in models belonging to the class (i), see Sec.\ref{sec-models}.

From now on, we focus on the case of D-term inflation. In this case, the
masses of the Higgs and gauge fields forming the string are
identical.
The string mass per-unit-length is then given by \cite{Mark}:
\begin{equation}
\mu  = 2 \pi \xi_x . \label{eq:Gmu}
\end{equation}
Hence cosmic strings forming at the end of inflation
are very heavy, and we expect them to contribute to the 
temperature anisotropies in the CMBR as well as to the fluctuations in 
energy density
of the universe which lead to structure formation. 
These cosmic strings may also have other cosmological
consequences as will be discussed in Sec.\ref{sec-models}. In the following 
section,
we will determine the relative contributions from inflation and cosmic strings 
to the CMBR.

\subsection{The predictions}
\label{sec-infl}

The temperature fluctuations in the CMBR are proportional to the density 
perturbations 
which were produced in the very early Universe and lead to structure 
formation: ${\delta T \over T} = {1\over 3}  {\delta \rho \over \rho }$.
In the case of the 
F-term and D-term hybrid inflationary scenarios discussed in this paper, 
cosmic strings form at the end of inflation. Thus we have contributions to
$\delta \rho \over \rho $ and hence to ${\delta T \over T}$ from both inflation 
and cosmic strings in different proportions which we estimate. 
Whether the strings are topologically
stable depends on the specific model which is considered. In the following,
we suppose the strings to be topologically stable.
We take the normalisation 
to COBE for inflation from Ref.\cite{BLW} and for cosmic strings 
from Refs.\cite{Paul}. 

It is usual to expand the temperature anisotropies in
spherical harmonics:
\begin{equation}
{\delta T \over T }(\theta ,\phi)  = \sum_{l,m} a_{lm} Y_{lm}
\end{equation}
and then to work with the multipole moments (or angular power spectrum):
\begin{equation}
C_l = {1 \over 2l+1} \sum_{m=-l}^{m=+l} |\! a_{lm}\! |^2 \label{eq:Clsum}
\end{equation}
which contain all information of the CMB (if it is Gaussian). 

The density perturbations (and other quantities) are usually expressed in
Fourier space, because the different fluctuation modes are uncorrelated 
and the solution is greatly simplified:
\begin{equation}
\delta(k) = \int d^3x {\delta \rho \over \rho}(x) e^{ik.x} .
\end{equation}
The spectrum of density perturbations is
usually assumed to be a power law:
\begin{equation}
\delta(k) \propto k^{n-1} ,
\end{equation}
as suggested by observations.
The exponent $n$ is called the spectral index.

The spectrum of density perturbations coming from inflation and the spectral
index can be
calculated analytically using the slow roll parameters.
These are given by \cite{LythLid}:
\begin{equation}
\epsilon = {M_{pl}^2 \over 16 \pi} \left({V'\over V}\right)^2 \hspace{1cm} 
{\rm and} \hspace{1cm}  \eta ={M_{pl}^2 \over 8 \pi} {V''\over V}
\end{equation}
where prime denotes derivative with respect to the inflaton field $|S|$.
The conditions for inflation to happen are \cite{LythLid}:
\begin{equation}
\epsilon \ll 1 \hspace{1cm} {\rm and} \hspace{1cm} | \eta | \ll 1 .
\end{equation}
For D-term inflation, with an effective potential given by
Eq.(\ref{eq:Dscalareff}), the slow parameters are:
\begin{equation}
\epsilon = {M_{pl}^2 g^4\over 128 \pi^2 } {1\over |S|^2}
\hspace{1cm} {\rm and} \hspace{1cm} 
\eta = - {M_{pl}^2 g^2\over 64 \pi^3} {1\over |S|^2}. \label{eq:eta}
\end{equation}
And we see that $\epsilon$ is always smaller than $\eta$. Consequently,
in this case inflation ends when $| \eta  | = 1$. The value
of the inflaton field at the end of inflation is:
\begin{equation}
S^2_{end} = {M^2_{pl} g^2\over 64 \pi^{3}}.
\end{equation}

The spectrum of density perturbations 
coming from
inflation is given by \cite{LythLid}:
\begin{equation}
\delta_H^2(k) = (1+4.0\epsilon - 2.1\eta) {512 \pi \over 75} 
{V^3 \over M_{pl}^6 V^{'2}}|_{k = a H} \label{eq:pert}
\end{equation}
where $k$ is the comoving wave number, $a$ is the cosmic scale factor and $H$
is the Hubble parameter. The right-hand side of Eq.(\ref{eq:pert}) must 
be evaluated at the epoch of horizon exit.
The spectral index of density perturbations can also be expressed in terms of 
the slow parameters:
\begin{equation}
n = 1 - 6 \epsilon + 2 \eta \label{eq:n}
\end{equation}
and can be evaluated at any scale. We also evaluate it at the epoch 
of horizon exit. The number of e-foldings between two 
values of the inflaton field $S_1$ and $S_2$ is given by 
\begin{equation}
N(|S|) \simeq {8\pi \over M_{pl}^2 }\int_{S_2}^{S_1} {V\over V'} d|S| .
\end{equation}
Assuming that it takes 60 e-folds for the cosmological scales 
to leave the horizon, the value of the inflaton field at the epoch 
of horizon exit is then: 
\begin{equation}
S^2_{60} = {121 g^2 \over 64 \pi^{3}} M_{pl}^2. \label{eq:S60}
\end{equation}
Using Eqs. ({\ref{eq:eta}), ({\ref{eq:n}) and
({\ref{eq:S60}) we find the value for the spectral index:
\begin{equation}
n = 0.98
\end{equation}
where we have used the relation ${g^2 \over 4\pi} = {1\over 25}$. 

For power law inflation, the multipole moments (\ref{eq:Clsum}) 
are related to spectrum of density perturbations at horizon exit 
(\ref{eq:pert}) by:
\begin{equation}
l(l+1) C_l^{infl} = f(n,l) \delta_H^2 \label{eq:Clinfl}
\end{equation}
where $f$ is a function of the spectral index $n$ and of the spherical
harmonic $l$ which is given by \cite{LythLid}:
\begin{equation}
f(n,l) = {\pi^{3\over 2} \over 4} l(l+1) {\Gamma({(3-n)\over 2}) 
 \Gamma(l+ {(n-1)\over 2}) \over \Gamma({(4-n)\over 2}) 
 \Gamma(l + {(5-n)\over 2})} .
\end{equation}

For cosmic string scenarios, the multipole moments are proportional
to the string mass per unit length $\mu = 2 \pi \xi_x$. Numerical simulations 
give, for $l \leq 20$, \cite{Paul}:
\begin{equation}
l(l+1) C_l^{str} \sim 350 (G\mu )^2 \label{eq:Clstr}
\end{equation}
where $G$ is Newton's constant.

With both inflation and cosmic strings the multipole moments
are given by given by:
\begin{equation}
C_l^{tot} = C_l^{infl} + C_l^{str} . \label{eq:Cl}
\end{equation}
The best fitting to COBE data occurs at the fourteenth multipole \cite{BLW}.
The normalisation to COBE for inflation yields \cite{BLW}:
\begin{equation}
\delta^{norm}_H (n,r)= N_{infl} = 1.91 \times 10^{-5} {\exp{(1.01(1-n))}\over 
\sqrt{1+0.75r}} = 1.94 \times 10^{-5} \label{eq:Binfl}
\end{equation}
for the density perturbations, where $r$ measures the relative importance 
of gravitational waves and density perturbations to the relevant 
multipole moment. Using the results of Ref.\cite{BLW}, we get $r = 2.7 \times
10^{-3}$, and hence the gravitational wave spectrum can be neglected here. 
Note
that this is a commun feature of all hybrid models. The uncertainty in 
$\delta^{norm}_H$ is 9\%. Now the normalisation to COBE for cosmic 
strings yields \cite{Paul}:
\begin{equation}
(G\mu )^{norm} = N_{str}  = 1.05^{+ 0.35}_{-0.20} \times 10^{-6} \label{eq:Nstr}
\end{equation}
for the string mass per unit length. Combining equations 
(\ref{eq:Clinfl}), (\ref{eq:Clstr}),(\ref{eq:Cl}),
(\ref{eq:Binfl}) and (\ref{eq:Nstr}) leads to a normalisation equation
for a mixed scenario with inflation and cosmic strings
\begin{equation}
1 = \left({G\mu \over N_{str}}\right)^2 + \left({\delta_H \over N_{infl}}
\right)^2 . \label{eq:norm}
\end{equation}
Using Eqs.(\ref{eq:Gmu}), (\ref{eq:pert}), (\ref{eq:Dscalareff}) and 
(\ref{eq:S60}), Eq.(\ref{eq:norm}) becomes:
\begin{equation}
1 = \left({\xi \over M_{pl}^2}\right)^2 \left({4 \pi^2 \over N_{str}^2} + 
{\alpha_{60} \, 256 \, \pi^2 \, 121 \over 75 \, N_{infl}^2}\right) 
\label{eq:norm1}
\end{equation}
where $\alpha_{60} = 1 + 4.0 \epsilon_{60} - 2.1 \eta_{60}$.

From Eq.(\ref{eq:norm1}), we find that the ${\rm U}(1)_x$ SSB 
scale is constraint be:
\begin{equation}
\xi_x^{1\over 2} =  4.7^{+0.5}_{-0.6} \times 10^{15} \, \, \, {\rm GeV}.
\end{equation}
From Eqs.(\ref{eq:Dscalareff}), we find that during inflation the 
Universe is dominated by the following energy density: 
\begin{equation}
V^{1\over 4}_0 \simeq 3.3 \times 10^{15} \, \, \, {\rm GeV} .
\end{equation}
Finally, from Eq.(\ref{eq:norm1})
we find that cosmic strings contribute to the $C_l$s at the level of:
\begin{equation}
75^{+10}_{-15} \, \, \, \% .
\end{equation} 
Note that it is in the case of D-term
inflation that the strings contribute the most to the CBR.

This result shows that the contribution from strings to the CBR temperature 
anisotropies and therefore to the density perturbations in the early
Universe which lead to structure formation is non negligible. 
Density perturbations due 
to a mixed scenarios 
with inflation and cosmic strings should be computed. The effects 
of both inflation and cosmic strings on the temperature of the CBR 
should be taken into account.  

The results obtained in this section 
are somewhat similar to those obtained in the case of F-term inflation
in both globally supersymmetric theories \cite{paper3}
locally supersymmetric ones  \cite{RiottoLinde}.

\section{The models}
\label{sec-models}

In this section, we discuss the three classes of models to which 
the construction discussed in this paper can be applied, 
see Sec.\ref{sec-building}. We take models which have an intermediate 
${\rm SU}(3)_c \times {\rm SU}(2)_L \times {\rm U}(1)_R \times {\rm U}(1)_{B-L}$ 
gauge symmetry
as illustration. These models are very interesting from both a particle physics
and a cosmological point of view.
 First, they satisfy all the requirement for successful inflation 
which can arise
without fine tuning. They also conserve R-parity automatically, predict both 
hot and cold dark-matter in the form of a massive neutrino and the
lightest-superparticle respectively, and lead to a scenario for
baryogenesis via leptogenesis. The minimal 
theory beyond the standard model which belongs to this class of models 
is the left-right symmetry 
$G = {\rm SU}(3)_c \times {\rm SU}(2)_L \times {\rm SU}(2)_R \times {\rm 
U}(1)_{B-L}$ and the
minimal grand unified theory is SO(10).

\subsection{Inflation from the unified theory itself, F-term inflation}
\label{sec-Fterm}

We consider here {\bf case (i)} where the ${\rm U}(1)_x$ symmetry is the 
subgroup of
a non abelian gauge group belonging to the visible sector.
The general SSB pattern which is discussed
is the following:
\begin{eqnarray}
G \times{\rm SUSY} &\stackrel{M_{GUT}}\rightarrow & ... \rightarrow H 
\times{\rm SUSY} \stackrel{M_{\rm x}}\rightarrow 
K \times{\rm SUSY} \rightarrow ... \nonumber \\
&\rightarrow & {\rm SU}(3)_c \times {\rm SU}(2)_L 
\times {\rm U}(1)_Y \times{\rm SUSY} \stackrel{M_{\rm Z}}\rightarrow {\rm 
SU}(3)_c \times 
{\rm U}(1)_Q \label{eq:SSB4}
\end{eqnarray}
where H is a subgroup of G which contains a ${\rm U}(1)_x$ gauge symmetry,
K is a subgroup of H which does not contain ${\rm U}(1)_x$. 
G can be identified with H and/or  K with the standard model gauge group ${\rm 
SU}(3)_c 
\times {\rm SU}(2)_L 
\times {\rm U}(1)_Y$, where appropriate. 

The ${\rm U}(1)_x$ 
symmetry is broken at $M_x$ by a pair of Higgs fields $\Phi_x$ and 
$\overline{\Phi}_x$ in complex conjugate
representations of G which transform non trivially under G. 
The ${\rm U}(1)_x$ symmetry is a subgroup of a non abelian gauge group, and 
hence
the hybrid 
inflationary scenario must be of the F-term type. 
The inflaton is a scalar field 
singlet under G which couples to the $\Phi_x$ and $\overline{\Phi}_x$
fields, with a superpotential in the 
inflaton sector given by Eq.(\ref{eq:Finfl}).
The phase transition $H \stackrel{M_x}\rightarrow K$ takes
place at the end of inflation, and cosmic strings form, see 
Sec.\ref{sec-strings}. 
The intermediate symmetry group H
(and K when $K\neq {\rm SU}(3)_c \times {\rm SU}(2)_L \times {\rm U}(1)_Y$) must 
be chosen such 
that the symmetry breaking
$H\rightarrow K$ (and $K\rightarrow {\rm SU}(3)_c \times {\rm SU}(2)_L \times 
{\rm U}(1)_Y$
respectively)
do not lead to the formation of monopoles or domain walls, which would 
make the model in conflict with the standard cosmology (unless this latter
symmetry breaking scale be very low). 

Let us now turn to models with an intermediate ${\rm U}(1)_{B-L}$ gauge 
symmetry. The
SSB pattern is now given by:
\begin{eqnarray}
G \times{\rm SUSY} &\stackrel{M_{GUT}}\rightarrow & ... \rightarrow 
 {\rm SU}(3)_c \times {\rm SU}(2)_L 
\times {\rm U}(1)_R \times {\rm U}(1)_{B-L} \times{\rm SUSY}  \nonumber \\
&\stackrel{M_{\rm B-L}} \rightarrow & {\rm SU}(3)_c \times {\rm SU}(2)_L 
\times {\rm U}(1)_Y \times{\rm SUSY} \stackrel{M_{\rm Z}}\rightarrow {\rm 
SU}(3)_c \times {\rm U}(1)_Q
\end{eqnarray}
where the unified gauge group G contains ${\rm U}(1)_{B-L}$. 
The minimal models are those where G is identified 
with ${\rm SU}(3)_c \times {\rm SU}(2)_L \times {\rm SU}(2)_R \times {\rm 
U}(1)_{B-L}$, and SO(10)
for semi-simple gauge groups,
which break directly down to ${\rm SU}(3)_c \times {\rm SU}(2)_L 
\times {\rm U}(1)_R \times {\rm U}(1)_{B-L}$. The study of 
${\rm SU}(3)_c \times {\rm SU}(2)_L \times {\rm U}(1)_R \times {\rm U}(1)_{B-L}$ 
on its own
is also possible.
We now couple the inflaton field with the pair of Higgs superfields
used to break ${\rm U}(1)_{B-L}$, namely $\Phi_{B-L}$ and 
$\overline{\Phi}_{B-L}$.

The superpotential is constructed by 
considering different sectors and by adding each sector as in 
Eq.(\ref{eq:super}).
The effective scalar potential can then be written as in 
Eq.(\ref{eq:scalar}).
The different sectors are the following:
\begin{itemize}
\item The inflationary sector, which is described by:
\begin{equation}
W_{infl}(S,\Phi_{B-L},\overline{\Phi}_{B-L}) = \alpha S \Phi_{B-L} 
\overline{\Phi}_{B-L} - \mu^2 S
\end{equation}
where $\Phi_{B-L}$ and $\overline{\Phi}_{B-L}$ are Higgs fields used
to break $B-L$, with opposite $B-L$ charges. They transform non trivially 
under G.
The components of $\Phi_{B-L}$ and $\overline{\Phi}_{B-L}$ which acquire a vev
transform as gauge singlet under the standard model gauge group. For SO(10) 
for example, $\Phi_{B-L}$ and 
$\overline{\Phi}_{B-L}$ can be a pair of $16 + \bar{16}$ dimensional Higgs 
representation or
$126 + \bar{126}$ dimensional one. 
$S$ is a chiral superfield whose bosonic component is a singlet 
under G which plays the role 
of the inflaton. $\alpha$ and $\mu$ are two positive constants. 
 $W_{infl}$ gives rise to a period of false vacuum hybrid inflation and 
${\rm U}(1)_{B-L}$ breaks spontaneously at the end of inflation with the fields 
$S$, $\Phi_{B-L}$ and 
$\overline{\Phi}_{B-L}$ acquiring vevs: $S = 0$, $\langle |\Phi_{B-L}| > =
\langle \overline{\Phi}_{B-L} \rangle = {\mu\over \sqrt{\alpha}}$.

\item The GUT sector, which is described by $W_{GUT}(A_i)$, $i=1...n$.
The $n$ Higgs 
fields $A$ are in various representations of G. $V_{GUT}$ must have a global
minimum such that $G$ is broken down to ${\rm SU}(3)_c \times {\rm SU}(2)_L 
\times 
{\rm U}(1)_R \times {\rm U}(1)_{B-L}$ with the $A$ fields acquiring vev. An 
example 
of $W_{GUT}$ in the case of SO(10) can be found in Refs.\cite{babu,paper3}.
In the case of SO(10), $W_{GUT}$ will involve Higgs superfields in the 
adjoint representation,
and possibly in the 54 dimensional representation.

\item The electroweak sector, which is described by $W_{ew}(H_1,H_2)$, breaks 
${\rm SU}(3)_c \times {\rm SU}(2)_L \times {\rm U}(1)_Y$ down to ${\rm SU}(3)_c 
\times {\rm U}(1)_Q$.
Hence the Higgs fields $H$ and $H'$ are in complex 
representations of G which must contain {\rm SU}(2) doublets. Extra coupling 
with the 
Higgs superfields of the GUT sector may be needed to solve the doublet-triplet
splitting problem; see for example Ref.\cite{babu}. 

\end{itemize}

Note that the inflationary and GUT sectors may have to be linked
in order to avoid any unwanted light pseudo-Goldstone particles. 
This can be done in a number of ways, 
for example by introducing extra Higgs fields chosen in such 
a way that they do not affect the vevs of the other Higgs, or by 
introducing non-renormalisable couplings between the $A$ and the $\Phi_{B-L}$
fields. This will not affect the dynamics of the model discussed
below.  Furthermore, as mentioned above, 
the electroweak 
sector will usually be facing the second hierarchy problem. And hence
$W_{ew}$ also involves a coupling with one of the A fields and $H_1$ and $H_2$
such that {\rm SU}(2) triplets get a large mass while the doublets remain light.

The scalar potential (\ref{eq:scalar}) is given by:
$$
V_{tot} = V_{GUT} (A_i) + V_{infl} + V_{ew}(H_1,H_2) + V_{soft}
$$
The usual scenario then applies \cite{DvaSha}. We impose chaotic initial 
conditions. The 
potential is very flat in the $|S|$ direction. The $A$ fields acquire a vacuum 
expectation 
value at a scale $\sim 2\times 10^{16}$ GeV, the vanishing conditions for the
F and D term must be satisfied, and $G$ breaks down to ${\rm SU}(3)_c \times 
{\rm SU}(2)_L \times 
{\rm U}(1)_R \times {\rm U}(1)_{B-L}$. Topologically stable monopoles
form. 

Initially
$|S| >> S_c = {\mu \over \sqrt{\alpha}}$, value of the inflaton field for which 
$V_{tot}$ as a local minimum
in the $\Phi_{B-L}$ and $\overline{\Phi}_{B-L}$ directions at 
$\Phi_{B-L} = \overline{\Phi}_{B-L} = 0$. So the $\Phi_{B-L}$ and 
$\overline{\Phi}_{B-L}$ fields settle down 
to this local minimum. There is a non-vanishing $F_S$ and
hence a non-vanishing vacuum energy density, 
$V_{eff}^{1\over 4} = \mu$,
the slow roll conditions are satisfied  and
inflation starts. There is a non-vanishing vacuum energy density and  
supersymmetry is broken. Quantum correction to the effective
potential can be taken into account and play a crucial role in
pulling the inflation field down the potential \cite{DvaSha}. When $|S|$ 
falls below $S_c$, the slow roll conditions are violated, and inflation stops.
The unwanted monopoles have been inflated away. At the
end of inflation
${\rm SU}(3)_c \times {\rm SU}(2)_L \times {\rm U}(1)_R \times {\rm U}(1)_{B-L}$ 
breaks down to
${\rm SU}(3)_c \times {\rm SU}(2)_L \times {\rm U}(1)_Y$ and $B-L$ cosmic 
strings form. 
Baryogenesis via leptogenesis takes place at the
end of inflation \cite{paper4}. In this type
of scenario, the $B-L$ breaking scale is constraint by COBE 
to be $M_{B-L}\sim 5 \times 10^{15}$ GeV, see Sec.\ref{sec-infl}. 
Also density perturbations in the early universe are 
due to a mixed scenario with inflation and cosmic strings,
with a bigger contribution from inflation, see Ref.\cite{paper3,RiottoLinde} and
Sec.\ref{sec-infl}.

\subsection{Inflation from an extra {\rm U}(1), D-term inflation}
\label{sec-Dterm}

If the ${\rm U}(1)_x$ symmetry is not
the subgroup of a non abelian gauge group, it can 
arise from both the non-vanishing vev of a F-term or that of a D-term,
in particular for
globally supersymmetric models. However, 
whereas in globally supersymmetric GUTs both F-term and D-term inflation
can occur, in supergravity GUTs D-term inflation is strongly favoured,
see Refs.\cite{Dvali} and Sec.\ref{sec-intro}. This however 
requires the presence of {\rm U}(1) gauge symmetry with $Tr Q \neq 0$.
We therefore consider here only the case of inflation driven 
by the non-vanishing vev of a D-term. We assume the existence of a 
Fayet-Iliopoulos
D-term for ${\rm U}(1)_x$.

The inflationary sector is now described by Eq.(\ref{eq:Dinfl}).
The only main change from F-term models comes from the 
superpotential in the inflationary sector. Hence switching from D-term to
F-term models in globally supersymmetric theories is straight forward.

\subsubsection{{\bf case (ii):} The extra {\rm U}(1) belongs to the visible 
sector}

The general SSB pattern which is assumed here is
the following:
\begin{eqnarray}
G \times {\rm U}(1)_{x} \times{\rm SUSY} & \stackrel{M_{GUT}}\rightarrow& ...
\rightarrow 
H \times {\rm U}(1)_{x} \times{\rm SUSY} \nonumber \\
&&\stackrel{M_{x}}\rightarrow 
 {\rm SU}(3)_c \times {\rm SU}(2)_L 
\times {\rm U}(1)_Y (\times{\rm SUSY}) \nonumber \\
 &&\stackrel{M_{\rm Z}}\rightarrow {\rm SU}(3)_c \times {\rm U}(1)_Q 
\end{eqnarray}
where rank(H) = 4. H is a subgroup of G and ${\rm SU}(3)_c \times {\rm SU}(2)_L 
\times {\rm U}(1)_Y$ is a subgroup of $H \times {\rm U}(1)_{x}$.
The scale $M_{GUT}$ is model depend but expected to be 
around $2 \times 10^{16}$ GeV
(it should be calculated by solving renormalisation group equations
for each specific model). 
The scale $M_{GUT}$ is constraint by the scale $M_x$ which is 
fixed by COBE.

The ${\rm U}(1)_x$ symmetry is broken by a pair of Higgs superfields 
$\Phi_x$ and $\overline{\Phi}_x$ which are charged under G. The
$\Phi_x$ and $\overline{\Phi}_x$ fields may transform non trivially under G.
The existence of a Fayet-Iliopoulos D-term associated ${\rm U}(1)_x$
is assumed. It sets the ${\rm U}(1)_x$ symmetry breaking scale.
The value of the Fayet-Iliopoulos term sets the ${\rm U}(1)_x$ symmetry breaking 
scale
and is constraint by COBE
data to be $\sim 5 \times 10^{15}$ GeV.

The phase transition associated with the spontaneous symmetry 
breaking of $H \times {\rm U}(1)_{x} \stackrel{M_x}\rightarrow 
 {\rm SU}(3)_c \times {\rm SU}(2)_L 
\times {\rm U}(1)_Y $ takes place at the end of inflation and
leads to the formation of cosmic strings with ${\rm U}(1)_x$ magnetic flux. 
Fermions
and bosons belonging to the visible sector may have non-vanishing $x$ charge.
This may have some cosmological effects. For example, surrounding $x$-charged
particles are subjected to the Aharonov-Bohm effect \cite{AB}. 
If fermions couple with the Higgs field $\Phi_x$ and acquire a mass
at $M_x$, there are fermion zero modes in the core 
of the strings. If these fermions are charged, the strings may carry
very large currents \cite{witten}. 

We take now again the examples of models where ${\rm U}(1)_x$ is 
identified with ${\rm U}(1)_{B-L}$. The scale $M_x$ is identified with
the $B-L$ breaking scale $M_{B-L}$. The existence of a 
Fayet-Iliopoulos term for ${\rm U}(1)_{B-L}$ is now assumed, and hence
${\rm U}(1)_{B-L}$ cannot be embedded in a larger gauge group
such as SO(10) here. The simplest model is given by the
SSB pattern:
\begin{eqnarray}
&&{\rm SU}(3)_c \times {\rm SU}(2)_L 
\times {\rm U}(1)_R \times {\rm U}(1)_{B-L} \times{\rm SUSY} \nonumber \\
&&\stackrel{M_{B-L}}
\rightarrow   {\rm SU}(3)_c \times {\rm SU}(2)_L 
\times {\rm U}(1)_Y (\times{\rm SUSY}) \nonumber \nonumber \\
 &&\stackrel{M_{\rm Z}}\rightarrow {\rm SU}(3)_c \times {\rm U}(1)_Q 
\end{eqnarray}
and the next to simplest one is given by the SSB pattern
\begin{eqnarray}
&&{\rm SU}(3)_c \times {\rm SU}(2)_L 
\times {\rm SU}(2)_R \times {\rm U}(1)_{B-L} \times{\rm SUSY} \nonumber  \\
&&\stackrel{M_{GUT}}\rightarrow  
{\rm SU}(3)_c \times {\rm SU}(2)_L 
\times {\rm U}(1)_R \times {\rm U}(1)_{B-L} \times{\rm SUSY} \nonumber \\
&&\stackrel{M_{B-L}}
\rightarrow   {\rm SU}(3)_c \times {\rm SU}(2)_L 
\times {\rm U}(1)_Y (\times{\rm SUSY}) \nonumber \\
 &&\stackrel{M_{\rm Z}}\rightarrow {\rm SU}(3)_c \times {\rm U}(1)_Q . 
\label{eq:LR}
\end{eqnarray}
Note that according to \cite{senjanovic} the Left-Right and $B-L$ breaking scale 
must be different.

The superpotential which implements the full spontaneous symmetry
breaking pattern from G down to ${\rm SU}(3) \times {\rm U}(1)_Q$ and gives
rise to a period of inflation can again be constructed 
by considering different sectors, and by adding the superpotentials
describing each sector. We consider the Left-Right model (\ref{eq:LR}).
The inflationary sector is now described by Eq.(\ref{eq:Dinfl}), with 
the $\Phi_x$ and $\overline{\Phi}_x$ fields identified with 
$\Phi_{B-L}$ and $\overline{\Phi}_{B-L}$. 
The existence of a Fayet-Iliopoulos D-term associated with the ${\rm 
U}(1)_{B-L}$
symmetry is assumed. The scalar potential in the inflaton sector
is given by Eq.(\ref{eq:Dscalar}). 

The scenario is the following.
We impose chaotic initial conditions. The initial value for the inflaton field 
is much greater than its critical value $S_c$, see Sec.\ref{sec-building}. Since 
the potential is flat in the $|S|$ direction, the potential can be minimised 
for fixed $|S|$ and the fields settle down to their local minimum. The 
$A$ fields acquire a non-vanishing vev and
${\rm SU}(3)_c \times {\rm SU}(2)_L \times {\rm SU}(2)_R \times {\rm 
U}(1)_{B-L}$G breaks down to 
${\rm SU}(3)_c \times {\rm SU}(2)_L \times {\rm U}(1)_R \times {\rm 
U}(1)_{B-L}$. Topologically
stable monopoles form.
The $\Phi_{B-L}$ and $\overline{\Phi}_{B-L}$ fields settle to zero. 
Once the $\Phi_{B-L}$ and $\overline{\Phi}_{B-L}$ are trapped in the local 
minimum
of potential, 
and inflationary period starts. Supersymmetry is broken and the one loop 
correction to the potential helps the inflaton field $|S|$ to slowly roll down 
the 
potential. The slow roll conditions are satisfied for
 $|S|^2 >> {g^2 \over  64 \pi^3} M^2_{pl} = S^2_{end}$. 
 When $|S|$ falls below $S_{end}$ inflation 
ends. The symmetry breaking ${\rm SU}(3)_c \times {\rm SU}(2)_L \times {\rm 
U}(1)_R \times {\rm U}(1)_{B-L}
\stackrel{M_{B-L}}\rightarrow {\rm SU}(3)_c \times {\rm SU}(2)_L 
\times {\rm U}(1)_Y$ takes place when $|S|$ falls below $S_c$. The unwanted 
monopoles have
been inflated away, and $B-L$ cosmic strings form. Baryogenesis
via leptogenesis takes place \cite{paper4}.
In this scenario, the $B-L$ breaking scale is constraint by COBE to be 
$ M_{B-L} \sim 5 \times 10^{15}$ GeV.

\subsubsection{{\bf case(iii):} The extra {\rm U}(1) belongs to the hidden 
sector}

The general SSB pattern is now given by:
\begin{eqnarray}
G \times [ {\rm U}(1)_{x} ] \times{\rm SUSY} &\stackrel{M_{GUT}}\rightarrow & 
...
H \times [ {\rm U}(1)_{x} ] \times{\rm SUSY} \nonumber \\
&&\stackrel{M_{x}}\rightarrow  {\rm SU}(3)_c \times {\rm SU}(2)_L 
\times {\rm U}(1)_Y (\times{\rm SUSY}) \nonumber \\
 &&\stackrel{M_{\rm Z}}\rightarrow {\rm SU}(3)_c \times {\rm U}(1)_Q 
\end{eqnarray}
where the ${\rm U}(1)_x$ symmetry now belongs to the hidden sector and breaks 
down to unity. H is a subgroup of G. 
The inflaton field is coupled with a pair of Higgs fields
$\Phi_x$ and $\overline{\Phi}_x$ used to break ${\rm U}(1)_x$. 
Effective field theories arising from
strings usually involve a certain number of {\rm U}(1) factors.
One of these {\rm U}(1)s could be the ${\rm U}(1)_x$ considered here. Note 
that it cannot be an anomalous {\rm U}(1) since the Fayet-Iliopoulos term which 
sets 
the scale of inflation must be $\xi_x \sim 5 \times 10^{15}$ GeV
as constrained by COBE, see Sec.\ref{sec-infl}, 
whereas for an anomalous {\rm U}(1) the Fayet-Iliopoulos term can be calculated 
by 
the Green-Schwarz mechanism, giving $\xi_{GS} = {{\rm Tr} (Q_x) g^2 
M^2 \over 192 \pi^2} >> \xi_x^{COBE}$.

All particles belonging to the visible sector are uncharged under this ${\rm 
U}(1)_x$
symmetry. And the $S$,$\Phi_x$ and $\overline{\Phi}_x$ are
 uncharged under G. The ${\rm U}(1)_x$
and the visible sector may only interact gravitaionally. 
The scale $M_{GUT}$ is model dependent but expected to be $\sim 2 \times 
10^{16}$ GeV
(should be calculated by solving renormalisation group equations). The 
scale $M_x$ is constraint by COBE to be $\sim 5 \times  10^{15}$ GeV.
The scale $M_{GUT}$ is independent of the scale $M_x$. 
The phase transition associated with the spontaneous symmetry 
breaking of ${\rm U}(1)_{x}$ down to unity 
takes place at the end of inflation and
leads to the formation of cosmic strings with ${\rm U}(1)_x$ magnetic flux.
These strings
interact with elementary  particles in the visible sector only gravitationally.
They can only be superconducting if there are fermions in the 
hidden sector acquiring mass with the field $\Phi_x$ acquiring a vev.

Let us now turn again back to our example, with a SSB pattern similar to that 
given by Eq.(\ref{eq:SSB4}), but the existence a ${\rm U}(1)_x$ gauge symmetry
in the hidden sector is now assumed. We now couple the inflaton field 
with the Higgs field used to break 
${\rm U}(1)_x$ rather than to $\Phi_{B-L}$. COBE sets the ${\rm U}(1)_x$ 
symmetry 
breaking scale at $\sim 5 \times 10^{15}$ GeV and
there is now considerable freedom in choosing the $B-L$ breaking scale and the 
Higggs potential in the $B-L$ sector.  Note that in the case of models 
where ${\rm U}(1)_{R}$ (${\rm U}(1)_{B-L}$) is the subgroup of a non abelian 
gauge group, topological stable
monopoles form when G breaks down to ${\rm SU}(3)_c \times {\rm SU}(2)_L 
\times {\rm U}(1)_R \times {\rm U}(1)_{B-L}$, and hence the $M_{GUT}$ scale must 
be greater 
than $M_x$.

\section{Conclusions}
\label{sec-concl}

False vacuum hybrid inflation emerges {\em naturally} in most
supersymmetric unified theories of the strong, weak and electromagnetic 
interactions with rank greater or equal to five. 
Inflation can a priori be driven by the non-vanishing vev 
of a F-term or that of a D-term. However, supergravity models favour
D-term inflation. On the other hand, D-term inflation requires the existence 
of a Fayet-Iliopoulos D-term which can only exist if the theory has
a {\rm U}(1) factor, and hence D-term inflation cannot arise from semi-simple 
gauge 
groups.

The models we have considered in this paper include a period of false vacuum
hybrid inflation with cosmic strings forming at the end of inflation. The 
scenario for
large scale structure formation implied by the models is a mixed scenario for
inflation and cosmic strings. In Sec.\ref{sec-strings} 
we made a good 
estimate to the relative contributions from inflation and cosmic strings to the
CMBR at the centre of the COBE data was made. We found that cosmic strings 
contribute
in this type of scenario at the level of 75 \% to the $C_l$'s. Hence their 
contribution 
is non negligible and should be taken into account when calculating the power 
spectrum or 
density perturbations in the early Universe which lead to structure formation.

Note that when cosmic strings form in the hidden sector they must interact 
only gravitationally with particles in the visible sector.

There is a particularly interesting class of models which fits in our 
discussion, those are models which involve an intermediate ${\rm SU}(3)_c 
\times {\rm SU}(2)_L \times {\rm U}(1)_R \times {\rm U}(1)_{B-L}$ gauge 
symmetry. 
Hybrid inflation
can arise in this type of scenario by coupling the inflaton field 
with the pair of Higgs fields used to break ${\rm U}(1)_{B-L}$. 
In that case $B-L$ cosmic strings form at the end of inflation and baryogenesis
via leptogenesis occurs at the end of inflation. The $B-L$ breaking scale
is also constrained by COBE to be $\sim 5 \times 10^{15}$ GeV.
If inflation comes form a hidden sector,
there is then considerable freedom for choosing the $B-L$ 
breaking scale. If another scale is involved in the model, say $M_{GUT}$, which
leads to the SSB $G \supset {\rm U}(1)_{R} ({\rm U}(1)_{B-L})$ 
down to ${\rm SU}(3)_c \times {\rm SU}(2)_L \times {\rm U}(1)_R \times {\rm 
U}(1)_{B-L}$, 
$M_{GUT}$ is then constrained to be greater than $\sim 5 \times 10^{15}$ GeV.

\section*{Acknowledgement}

The Author would like to thank D. Comelli, M. Hindmarsh, 
A. Liddle and D.Lyth
for very useful discussions. This work was supported by PPARC grant
no GR/K55967.

\end{document}